\documentclass{article}

\pdfoutput=1
\usepackage{arxiv}

\usepackage[utf8]{inputenc} 
\usepackage[T1]{fontenc}    
\usepackage{hyperref}       
\usepackage{url}            
\usepackage{booktabs}       
\usepackage{amsfonts}       
\usepackage{nicefrac}       
\usepackage{microtype}      
\usepackage{lipsum}
\usepackage{graphicx}
\graphicspath{ {./images/} }
\usepackage{hyperref}
\usepackage{amsmath}
\usepackage{bm}
\usepackage{siunitx}
 \DeclareSIUnit\pN{\pico\newton}
 \DeclareSIUnit[per-mode=symbol]\D{\um\squared\per\second}
 \DeclareSIUnit[inter-unit-product=\ensuremath{{}\cdot{}}]\pNnm{\pico\newton\nano\meter}
 \DeclareSIUnit\Molar{\textsc{m}}

\newcommand*{\kT}{k_{\textup{B}}\textup{T}}

\title{Inferring interaction potentials from stochastic particle trajectories}

\author{Ella M. King \\
 Physics Department, New York University \\
Simons Junior Fellow, Simons Foundation \\
\texttt{ellaking@nyu.edu} \\
\And
Megan C. Engel \\
 Department of Biological Sciences, University of Calgary,\\
 School of Engineering and Applied Sciences\\
 Harvard University\\ 
  \texttt{megan.engel@ucalgary.ca} \\
   \And
Caroline Martin, Alp M. Sunol, Qian-Ze Zhu \\%
 School of Engineering and Applied Sciences\\
 Harvard University\\ 
 \And
Sam S. Schoenholz \\
 OpenAI \\
\And 
Vinothan N. Manoharan and Michael P. Brenner \\ 
School of Engineering and Applied Sciences, Harvard University, \\
Physics Department, Harvard University \\
}

\begin{document}
\maketitle
\begin{abstract}
Accurate interaction potentials between microscopic components such as colloidal particles or cells are crucial to understanding a range of processes, including colloidal crystallization, bacterial colony formation, and cancer metastasis. Even in systems where the precise interaction mechanisms are unknown, effective interactions can be measured to inform simulation and design. However, these measurements are difficult and time-intensive, and often require conditions that are drastically different from \textit{in situ} conditions of the system of interest. 
Moreover, existing methods of measuring interparticle potentials rely on constraining a small number of particles at equilibrium, placing limits on which interactions can be measured. 
We introduce a method for inferring interaction potentials directly from trajectory data of interacting particles. We explicitly solve the equations of motion to find a form of the potential that maximizes the probability of observing a known trajectory. Our method is valid for systems both in and out of equilibrium, is well-suited to large numbers of particles interacting in typical system conditions, and does not assume a functional form of the interaction potential. We apply our method to infer the interactions of colloidal spheres from experimental data, successfully extracting the range and strength of a depletion interaction from the motion of the particles.
\end{abstract}


\section{Introduction}
Measuring interparticle interaction potentials reveals new physics. 
Such measurements have led to the discovery of long-range interactions in colloidal spheres with depletant particles~\cite{crocker1999entropic} and have been used to understand the unexpected like-charge attraction of colloidal spheres, attributed to many-body interactions in colloidal crystallites, hydrodynamic interactions with a surface, or interfacial solvation effects~\cite{larsen1997like, squires2000like, kubincova_interfacial_2020}. 
The direct measurement of off-target binding has led to improved monoclonal antibody selection for therapeutic applications~\cite{geng2014improving}. 
In addition to uncovering new physical principles, accurately measuring interaction potentials may provide insights into complex biological systems. Intricate cell-cell communications mediate cell processes ranging from cancer metastasis to tissue morphogenesis~\cite{brucknerLearningDynamicsCell2021}, while interactions between agents govern emergent behavior in active matter systems ranging from swarms of birds to colonies of bacteria~\cite{Bowick2022ActiveMatter}. 

Interaction potentials also provide a coarse-grained lens through which we can interpret and model complex system dynamics. 
For example, while measurements of cell-cell interactions at the molecular scale have produced profound insights into cellular adhesion mechanisms and immune response~\cite{bechtel2021strategies}, 
similar understanding of larger scale biological processes such as wound healing has been inaccessible.
While fully analytical models of such complex biological systems are difficult to realize, it is possible to instead integrate theoretical models with interactions directly inferred from experimental data, providing a framework for experimentally reproducible simulation with significantly reduced need for costly experiments. 
Further, accurate descriptions of these mesoscale interactions would open the door to the computational design of systems with bio-compatible components or bio-inspired active materials~\cite{bowick2022symmetry}.

In addition to improving modeling and computation, accurate knowledge of particle potentials can also be helpful for predicting dynamics. Integrating accurately measured interaction potentials into particle tracking has been shown to dramatically increase accuracy in colloidal systems, making it possible to infer trajectories even for highly correlated systems and expanding the range of experiments that can be performed~\cite{king2022correlation}. 

Current methods for experimental measurements of interaction potentials typically rely on observing small numbers of highly constrained particles. These particles are confined in optical traps to move in a single dimension, adhered to the wall of a flow cell, or constrained to the tip of a force sensor ~\cite{ducker_direct_1991, crocker1994microscopic, piech_prediction_2002, rogers_direct_2011, lee_optical_2016}. These methods for measuring interactions require complex experimental setups with conditions that may differ significantly from \textit{in situ} environments of the systems of interest. Even in methods where the particles are free to move, the analysis of the resulting particle trajectories typically assumes equilibrium motion. Nonequilibrium systems, including active matter and systems with many-body or velocity-dependent interactions, are inaccessible to these methods.

An alternative approach is the data-driven inference of force fields using experimental trajectories. This method has been applied to infer deterministic physical laws~\cite{lemos2022rediscovering, batzner20223, danielsAutomatedAdaptiveInference2015}, but there is a relative dearth of studies treating stochastic motion~\cite{Gao_2023}. Studies that have considered stochastic dynamics have either been restricted to single-~\cite{ beheiryInferenceMAPMappingSinglemolecule2015a,brucknerInferringDynamicsUnderdamped2020b} or two-particle~\cite{brucknerLearningDynamicsCell2021} dynamics; require assuming a functional form for the force~\cite{perezgarciaHighperformanceReconstructionMicroscopic2018a, sarfatiMaximumLikelihoodEstimations2017}; or rely on a complicated decomposition of force field onto a set of basis functions chosen \textit{a priori}, whose functional forms must be compatible with the inferred interparticle forces~\cite{brucknerInferringDynamicsUnderdamped2020b,frishman2020learning}. 

One strategy for combating the bias inherent in preselecting functional forms is to instead fit the potential to a general functional form. Graph Neural Networks (GNNs) are an especially appealing choice, as they implicitly include physical priors of locality and distance dependence of interparticle potentials. Fitting GNNs to simulations has been highly successful and is  well-established for deterministic systems~\cite{lemos2022rediscovering, batzner20223}. However, to our knowledge, no comprehensive treatment of fitting stochastic dynamics to GNNs has yet been attempted. Further, previous efforts to infer interaction potentials from trajectory data have primarily considered simulated data, while studies that have attempted experimental validation exclusively consider one-~\cite{beheiryInferenceMAPMappingSinglemolecule2015a,brucknerInferringDynamicsUnderdamped2020b} or two-particle~\cite{brucknerLearningDynamicsCell2021} experiments. 

Here, we present a maximum-likelihood-based approach for inferring interparticle interaction potentials directly from bulk particle trajectory data. These effective potentials are defined to be the best fit to trajectory data under the assumption that the system is well-described by a known equation of motion. We consider both overdamped Langevin (Brownian) dynamics and underdamped Langevin dynamics. We illustrate our approach for a variety of stochastic dynamics by inferring the parameters of a known functional form for the effective pair interaction potential, and then by inferring an arbitrary functional form with a GNN. We validate our method using experimental data of colloidal particles experiencing depletion interactions. Using a GNN, we reconstruct a pair interaction potential from experimental particle trajectories and validate the inferred parameters with prior characterizations of the system.

Our method is valid for any system for which the equations of motion are known. The method is equally applicable to both equilibrium and non-equilibrium data and to bulk and few-particle systems, significantly broadening the scope of systems for which interactions can be inferred. Experimental bulk trajectory data is abundant and relatively simple to obtain \cite{crocker_methods_1996, valentine_investigating_2001, prasad_confocal_2007}. We therefore expect our method to be of great utility to the experimental community, offering a straightforward means of characterizing interparticle interactions.

\section{\label{sec:params} Known functional form}
We introduce our method in the context of inferring parameters of an interaction potential with a known functional form. We focus on inferring the parameters of a Morse potential
\begin{equation}\label{eq:morse}
U(r) = \varepsilon (1 - \exp^{-\alpha (r - \sigma)})^2,
\end{equation}
where $\varepsilon$ is the depth of the potential well, $\alpha$ is the interaction range, and $\sigma$ is the particle size.

Our general approach is based on a maximum-likelihood formulation: under given dynamics, we identify the transition probability $P(\mathbf{r}_{i+1} | \mathbf{r}_{i})$ for a set of particles moving from positions $\mathbf{r}_{i}$ at frame $i$ to positions $\mathbf{r}_{i+1}$ at frame $i+1$. This probability is a function of the unknown pair potential $U$. We then compute the probability of observing the entire particle trajectory $\textbf{R(t)}$ consisting of $f$ frames by taking the product of transition probabilities for each step, given by
\begin{equation}
P(\mathbf{R}(t)) = \prod_{i=0}^{f-1} P(\mathbf{r}_{i+1} | \mathbf{r}_{i}).
\label{prob_product}
\end{equation}
To construct trajectories, we perform simulations of ensembles of particles in bulk. Given a proposed pair potential $U$, we compute the negative log likelihood of observing those trajectories and average across all sequential pairs of frames. We then find the potential $U$ that minimizes this average. 
\begin{figure*}[tpb]
	\centering
	\includegraphics[width=\linewidth]{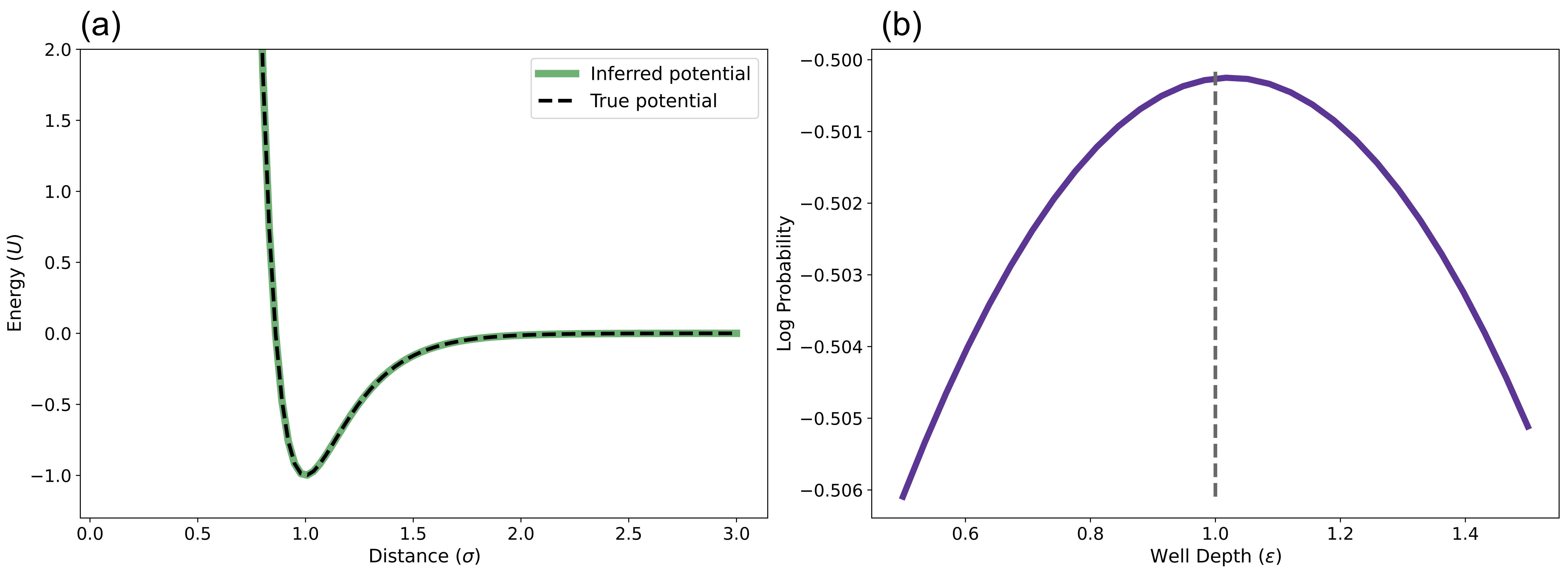}
 \caption{\label{fig:brownian_known} Inferring parameters of a known potential for simulated particles undergoing Brownian motion. (a) True interaction potential is shown with a black dashed line and inferred interaction potential is shown with a green solid line. The three parameters of the interaction are inferred simultaneously. (b) Log probabilities of observing the true trajectory for varying values for the depth parameter, $\varepsilon$. The remaining two parameters are set to the true values. Gray dashed line indicates the position of the value of the true value for $\varepsilon$.}
\end{figure*}
\subsection{Brownian Motion \label{sec:brownian}}
We begin by considering Brownian dynamics, which obey the overdamped Langevin equation
\begin{equation}\label{eq:brownian}
\mathbf{\dot r} = -\frac{1}{\gamma}\nabla U(\textbf{r}) +
\sqrt{\frac{2k_B T}{\gamma}}
\mathbf{\xi}(t),
\end{equation}
where $\mathbf{\dot r}$ is the time derivative of position, $\gamma$ is the friction coefficient, $U$ is the total interaction potential, $k_B$ is the Boltzmann constant, and $T$ is temperature. We assume that the total interaction potential $U(\textbf{r})$ can be written as a sum of pairwise interactions. We denote the sum over all pair interactions in frame $i$ as $U(\textbf{r}_i).$ $\mathbf{\xi}(t)$ describes the stochastic thermal motion in the system, and is given by delta-correlated Gaussian noise with mean zero: $\langle \mathbf{\xi}(t)\rangle = 0$, $\langle\mathbf{\xi}(t)\mathbf{\xi}(t')\rangle = \delta(t-t')$. These equations of motion assume no velocity-dependent effects, such as hydrodynamic interactions. While we expect we may be able to neglect these effects for many experimental systems, we also expect some amount of hydrodynamic coupling to contribute to the effective particle interactions. 

Given our knowledge of the noise statistics and dynamical equation, we assign the transition probability to any particular change in particle position $\mathbf{r}_i \rightarrow \mathbf{r}_{i+1}$ in a time interval $\Delta t$ as 
\begin{equation}
P(\mathbf{r}_{i+1} | \mathbf{r}_{i}) = \left(\frac{\gamma}{4\pi k_B T \Delta t}\right)^{\frac{d N }{2}}
e^{
\frac{-\gamma (\mathbf{r}_{i+1} - (\mathbf{r}_i - \frac{\Delta t}{\gamma} \nabla U(\textbf{r}_i)))^2}
{4k_B T \Delta t}
}.
\label{eq:prob}
\end{equation}
In the above equation, $N$ is the total number of particles, $d$ is the dimensionality of the position vector.

Substituting equation~\ref{eq:prob} into equation~\ref{prob_product} yields the total probability of observing a particular trajectory evaluated for all pairs of sequential frames in the trajectory. 
Assuming that the temperature ($T$), time between frames ($\Delta t$), and friction coefficient $\gamma$ are known, we can solve for the potential that maximizes the log likelihood averaged over all particle trajectories in our simulation.

We first consider a simulation of particles interacting via a Morse potential. We seek to find the set of parameters $\varepsilon$, $\alpha$ and $\sigma$ of the potential that maximize the probability of observing the set of trajectories. To find the most probable parameters, we perform a grid search over the varying values of $\varepsilon$, $\alpha$ and $\sigma$. For each point in the grid, we compute the negative log likelihood averaged over a set of 128 simulated Brownian particle trajectories of 500 simulation steps. The simulations are performed in JAX-MD, a molecular dynamics engine (see Appendix for further details). 

We find that we are able to recover the correct parameters of the interaction potential exactly for the case of simulated Brownian motion (Fig~\ref{fig:brownian_known}), with an inferred potential nearly identical to the simulated potential. As a representative example, we show in Fig~\ref{fig:brownian_known} the clear peak in the log probability of observing a value of $\varepsilon=1.01$, while the true value is 1.0. The remaining two parameters are similarly recovered, as shown in Fig~\ref{fig:brownian_known}. The likelihood we use here considers only Brownian noise. While this is valid for simulated data, true experimental data will contain additional noise sources, such as uncertainty in particle positions, that we neglect here. However, as we show in Section~\ref{sec:experiments}, this approach generalizes to experimental data despite this added uncertainty.

\subsection{Langevin Motion \label{sec:langevin_motion}}
We now consider the more general case of underdamped Langevin dynamics, for which the equation of motion is
\begin{equation}\label{eq:langevin}
m\mathbf{\ddot r} = -\nabla U(\textbf{r}) -\gamma \mathbf{\dot r} + \sqrt{2 \gamma k_B T}\xi(t),
\end{equation}
where $\textbf{r}$ is the vector of particle positions, $\mathbf{\dot r}$ is the vector of velocities, $\mathbf{\ddot r}$ is acceleration, $m$ is the mass, $\gamma$ is the friction coefficient, $U(\textbf{r})$ is the total interaction potential, $k_B$ is the Boltzmann constant, and $T$ is temperature. We assume that $U(\textbf{r})$ can be written as a sum over the interactions betwen pairs of particles in the system.
$\mathbf{\xi}(t)$ describes the stochastic thermal noise in the system, and is given by delta-correlated Gaussian noise with mean zero.

Extracting the transition probability from frame $i$ to frame $i+1$, $P(\mathbf{r}_{i+1}|\mathbf{r}_{i})$, is more subtle in the case of underdamped Langevin dynamics because it is a second-order differential equation in $\mathbf{r}$. The equation governing the evolution of a Langevin particle's phase space probability function, the Klein-Kramers equation, does not have a general analytic solution when the interaction potential $U(x)$ is non-zero.

The problem of accurately simulating underdamped Langevin motion was addressed by the development of integrators that split equation~\ref{eq:langevin} into deterministic and stochastic components and then combine the respective solutions. We base our transition probability estimator on one such integration scheme, a symplectic molecular dynamics integrator that uses the following splitting~\cite{leimkuhlerRationalConstructionStochastic2013,davidchack2015new}:

\noindent 1. Update momenta with increment $\Delta t/2$;\\
2. Update positions with increment $\Delta t/2$;\\
3. Perform a full stochastic step with increment $\Delta t$;\\
4. Update positions with increment $\Delta t/2$;\\
5. Update momenta with increment $\Delta t/2$. \\
This scheme is commonly called the ``BAOAB" Langevin splitting. Given the half-stepped momentum $p_0 \rightarrow p_1$, the stochastic step draws a new momentum from the distribution
\begin{align}
p_{new} &= \mathcal{N}(c_1 p_1, c_2^2 m) \label{eq:langevin_prob},
\end{align}
where $c_1 = \exp^{-\gamma \Delta t}$ and $c_2 = \sqrt{k_B T (1 - c_1^2)}$.
Given a trajectory and a proposed interaction potential, we invert this process until reaching the stochastic step. We begin with a set of four consecutive frames. We estimate the momentum in the previous ($P_{prev}$) and current ($P$) step using finite differences in the Stratonovich convention. We perform a half step backwards from $P$ using the inverted momentum update rule and a half step forwards from $P_{prev}$. We then compute the distribution given in Equation~\ref{eq:langevin_prob} using the updated $P_{prev}$ and apply it to the updated $P$ to estimate the probability of an observed change in momentum.

To ensure that the inference does not depend on the choice of integrator, we simulate using both the ``BAOAB'' Langevin integrator and a simpler, second order integrator of the Langevin equation~\cite{vanden-eijndenSecondorderIntegratorsLangevin2006}. Both sets of simulations yield the same results. 
\begin{figure}[tpb]
	\centering
	\includegraphics[width=\linewidth]{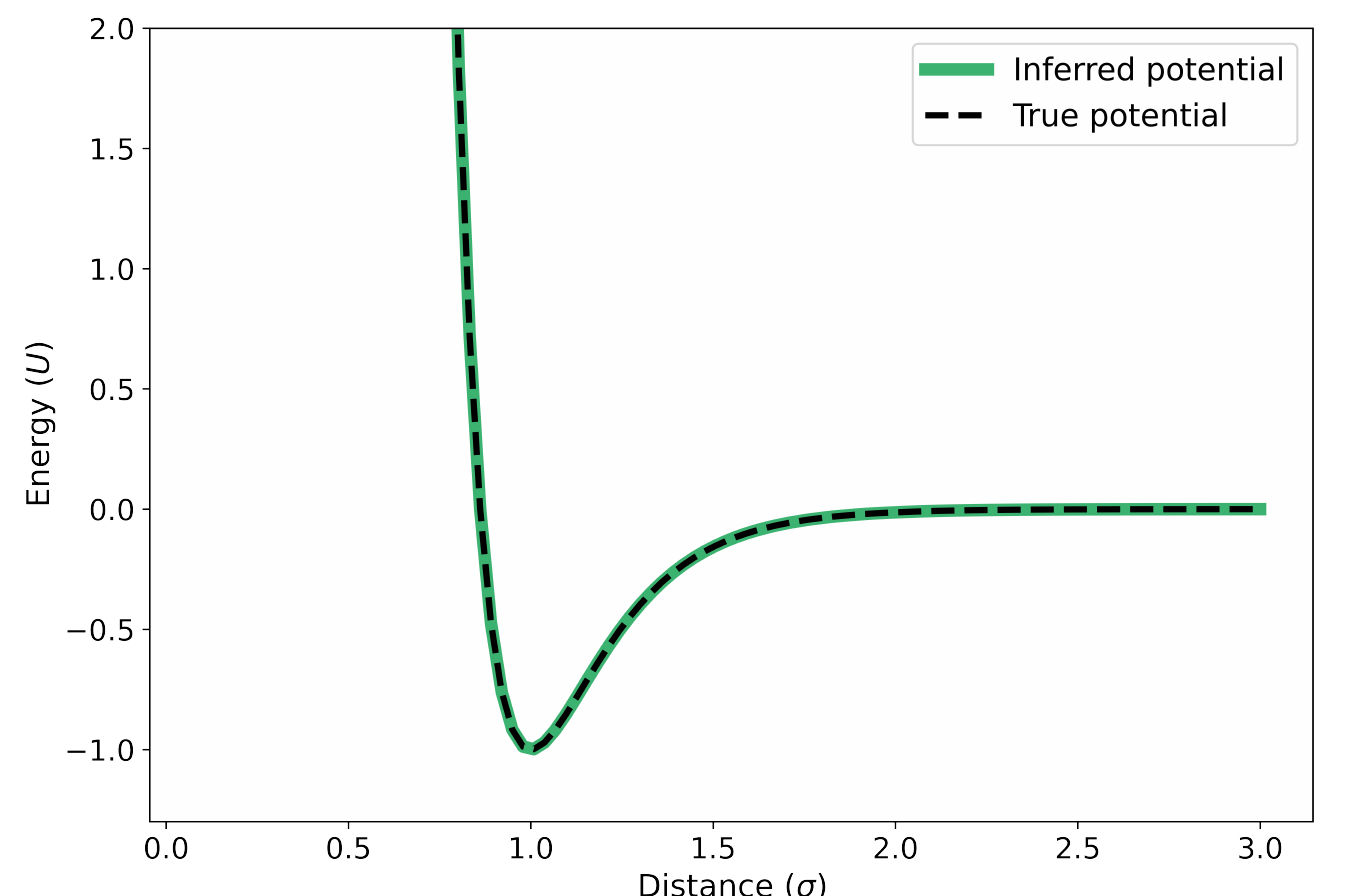}
	\caption{\label{fig:langevin_known} Inferring parameters of a known potential for particles undergoing Langevin motion. The true interaction potential is shown with a black dashed line and the inferred interaction potential is shown with a green solid line. The three parameters of the interaction are inferred simultaneously.}
\end{figure}

We are able to accurately reconstruct the true interaction potential from bulk particle trajectories undergoing underdamped Langevin motion (Fig~\ref{fig:langevin_known}). In this case, however, we require 5,000 data points to recover parameters with reasonable accuracy, compared to the only 500 required for particles undergoing Brownian dynamics. We attribute this increase to errors introduced by the finite difference estimates of the momentum, which also leads to a slight decrease in accuracy of the inferred parameters compared to the Brownian case. We estimate $\varepsilon = 0.9$ (true value of 1.0), $\alpha = 5.3$ (true value of 5.0), and $\sigma=1.0$ (true value of 1.0). Despite these discrepancies in values, we find good agreement in the functional forms of the potentials (Fig~\ref{fig:langevin_known}), 
indicating that there are covariances among the three parameter estimates.

\section{\label{sec:neural_net} Arbitrary functional form}
While inferring parameters of a known potential can provide rich information about a system of interacting particles, assuming a specific functional form limits the amount of information that can be extracted. Inferring an interaction potential with an arbitrary functional form broadens the method to more complex interaction potentials and allows the method to reveal potentially unexpected forms.

Here, we parameterize an arbitrary potential using a neural network. Specifically, we use NequIP, an $E(3)$ equivariant neural network~\cite{batzner20223} that was designed to find molecular dynamics potentials that approximate density functional theory (DFT) simulations. This neural network can capture a wide range of functional forms, including long-range interactions. While we only consider identical particles with pair potentials, this approach can easily treat particles with additional features, such as orientations, or many-body interactions. 
Our method can readily accommodate different neural network architectures, making our setup simple to integrate with rapidly evolving neural network architectures.
Previous work employing GNNs used alignment with deterministic trajectories as a cost function~\cite{lemos2022rediscovering}, but this strategy is inaccessible to stochastic motion. Instead, we leverage the same maximum likelihood approach we introduce in Section~\ref{sec:params}. The GNN input is a set of particle positions in a single frame and its output is the total potential energy of that frame. The cost function used to train the network parameters is the negative log likelihood of obtaining the observed particle motion between pairs of consecutive frames. We average the log probability over a subset of such pairs, constituting a training batch, to update the GNN parameters.

\begin{figure}[h!tpb]
	\centering
	\includegraphics[width=\linewidth]{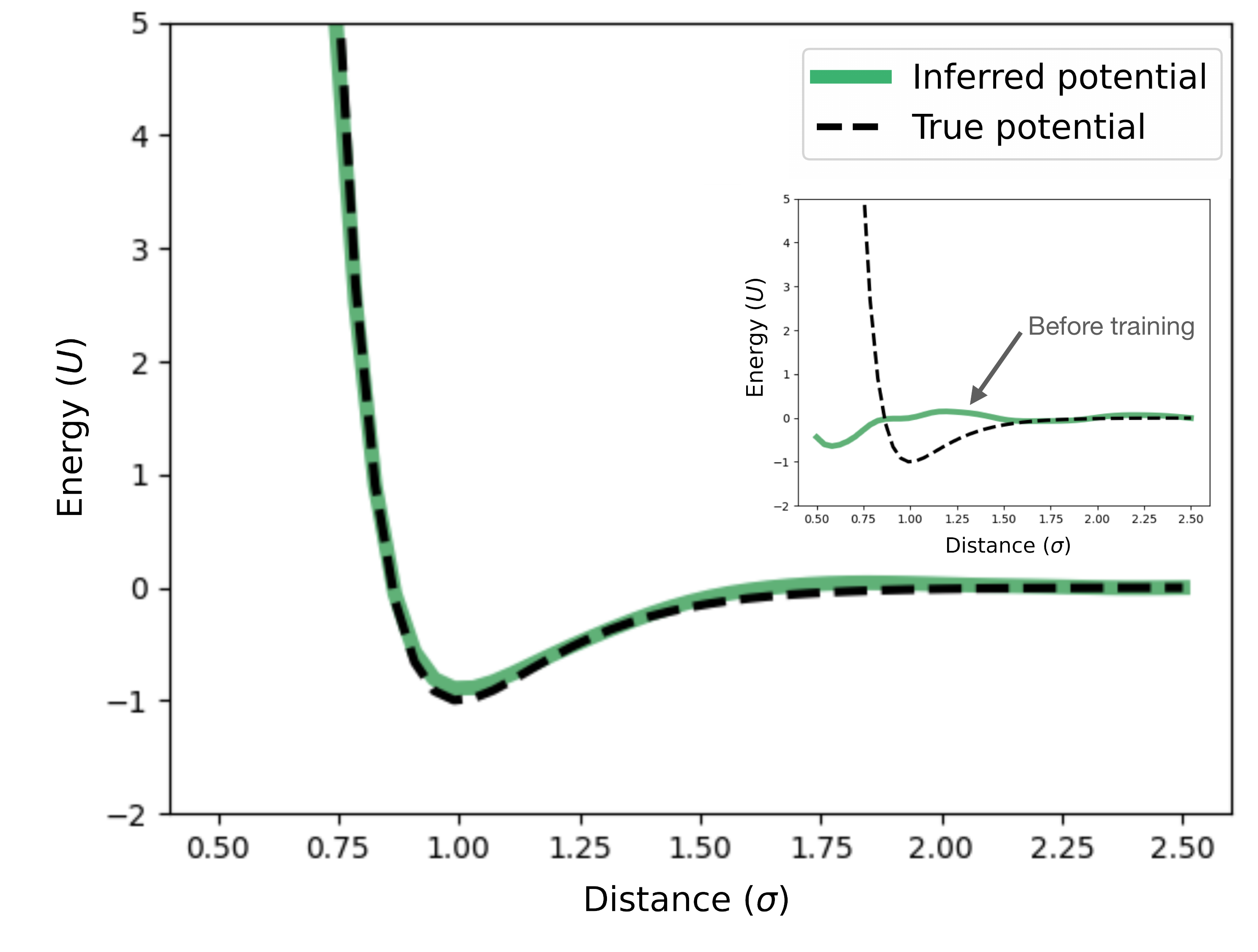}
	\caption{\label{fig:brownian_arbitrary} Inferring an arbitrary potential for simulated particles undergoing Brownian motion. The true interaction potential is shown with a black dashed line and the inferred interaction potential is shown with a green solid line. The inset shows a sample form of the potential before training with randomized weights. No assumptions about the form of the potential were built into the model.}
\end{figure}

To visualize the resulting pair interaction potential, we evaluate the trained GNN on a system of two particles and obtain potential energies at a series of interparticle distances. We note that the network is trained on bulk systems of hundreds of particles, meaning that we are testing in a very different regime than we trained in. Nevertheless, applying this framework to the Brownian particle simulations described in section~\ref{sec:brownian}, we are able to accurately reconstruct the Morse interaction potential detailed above, as shown in Figure~\ref{fig:brownian_arbitrary}.

\section{\label{sec:experiments} Experimental Validation}
\begin{figure}[tpb]
	\centering
	\includegraphics{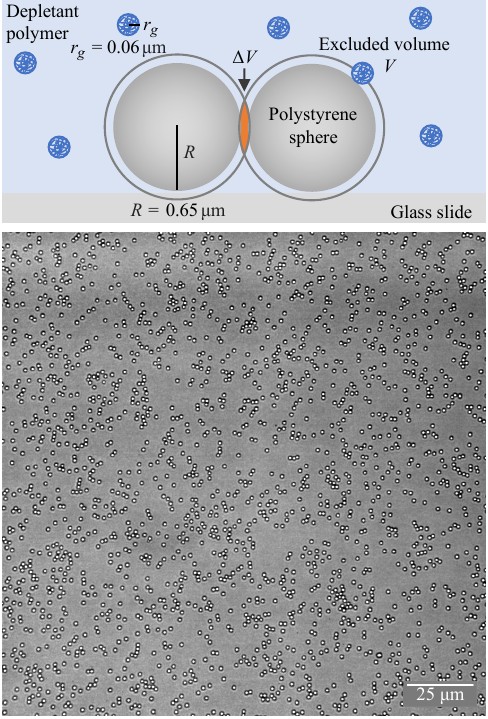}
	\caption{\label{fig:diagram} Diagram of experimental system. Charge-stabilized polystyrene spheres are suspended with depletant polymers with a radius of gyration of approximately \SI{0.06}{\micro\metre}. The depletants induce an entropic attraction between the larger spheres, with a range set by the size of the polymers. The spheres are depleted to a glass slide and freely diffuse in the quasi-2D sample chamber.}
\end{figure}

Thus far, we have demonstrated our method on simulated data. We now validate our approach on an experimental data set. We suspend \SI{1.3}{\micro\metre} charge-stabilized colloidal spheres in a solution of carboxymethyl cellulose polymers (radius of gyration approximately \SI{60}{\nano\metre}~\cite{hoogendam_persistence_1998, mondal_cooperative_2020}) in deionized water. The colloidal particles repel one another electrostatically in the absence of the polymer. Adding the polymer induces an entropic attraction, called the depletion interaction, that favors minimizing the volume excluded to the polymer (Fig~\ref{fig:diagram}). The overall interaction is the combination of electrostatic, van der Waals, and depletion interactions. The range of the depletion interaction is on the order of the size of the polymer, and the potential goes to infinity below contact (\SI{1.3}{\micro\meter} center-to-center distance) because the particles cannot overlap~\cite{AOpotential1954,maoDepletionForceColloidal1995}.

We allow the spheres to bind to the bottom surface of the chamber, confining them to a a quasi-2D geometry, and record bright-field microscope images as they diffuse and interact. We record short, high-frame-rate videos separated by longer time periods to ensure that the sampled trajectories are uncorrelated (Fig~\ref{fig:diagram}, see Appendix for additional experimental detail). We then use \texttt{trackpy} \cite{trackpy} to localize the colloidal particles to sub-pixel precision and link these positions across frames to determine trajectories. We discard trajectories of particles that leave the field of view, as well as the few particles that are outside the expected size of the colloidal particles, which we consider to be contaminants. 

\begin{figure*}[h!tpb]
	\centering
	\includegraphics[width=\linewidth]
 {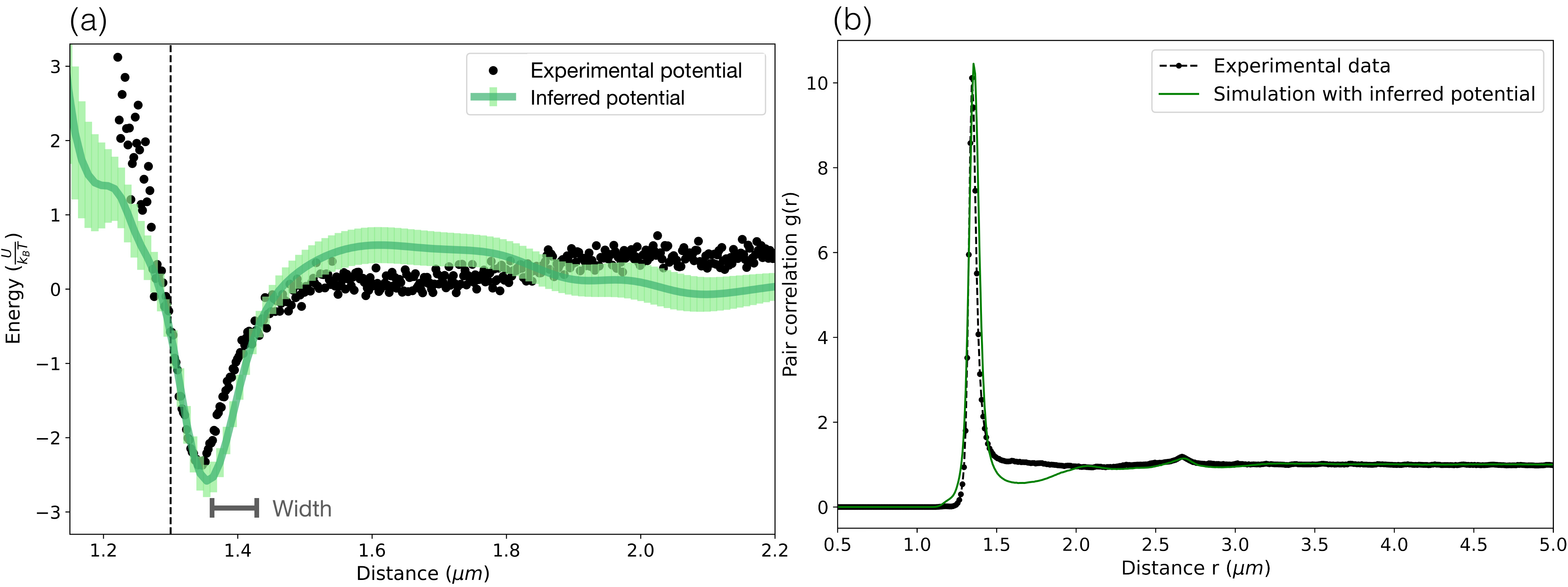}
	\caption{\label{fig:exp} Inferring an arbitrary interaction potential for experimental colloidal data with known interactions. (a) Inferred interaction potential is shown with a solid green line. The error bars show varying potentials found by running the training procedure with different random keys. Experimental data are shown with black scatter points. The vertical dashed line indicates the expected particle diameter. The estimate of half of the width of the potential is given by the horizontal gray bar. (b) Pair correlation function, $g(r)$. Experimental $g(r)$ is shown in black. To compute the inferred $g(r)$, we run a simulation with a representative interaction potential learned via the GNN. The $g(r)$ computed from the simulation data is shown in green.
 }
\end{figure*}

We train our GNN using 8,340 experimental snapshots collected from 60 distinct trajectories and infer the interaction potential shown in Fig~\ref{fig:exp}, assuming overdamped Langevin (Brownian) dynamics to analyze the data. We use the maximum likelihood method as described in section~\ref{sec:brownian} to infer the potential. While we account for Brownian noise in our likelihood estimate, we neglect additional experimental noise sources such as uncertainty in particle positions.

We find a minimum of the potential at \SI{1.33}{\micro\meter}. The Asakura-Oosawa model~\cite{AOpotential1954}, which assumes that depletent and colloidal particles are hard spheres, predicts a minimum at the particle diameter (here, approximately \SI{1.30}{\micro\meter}). This shift in the minimum likely arises because the particles are not hard spheres, and they electrostatically repel one another. 
The range of the interaction can be read off the potential as twice the width of the well from the minimum to the nearest region at which the potential is zero: \SI{0.16}{\micro\meter}, as shown in Fig~\ref{fig:exp}a.  This value is larger than the value predicted by the Asakura-Oosawa model, which predicts the range to be equal to the hard-sphere diameter of the depletant particles (here, the radius of gyration is approximately \SI{0.06}{\micro\meter}). The increase in the effective depletant size may also be explained by electrostatic effects, since the depletants are charged polymers. 

To further validate the inferred potential, we compare it to the potential inferred by assuming that system is in equilibrium and at low density, such that 
\begin{equation}
U(x) = -\kT \log{g(\textbf{r})}
\label{eq:pmf}
\end{equation}
where g(\textbf{r}) is the pair correlation function for a set of particle positions \textbf{r}.
The results are shown as black dots in Fig.~\ref{fig:exp}a. Equation~\ref{eq:pmf} yields the potential of mean force (PMF); however, at low particle densities, it can serve as a rough estimate of the pair potential, assuming that the system is at equilibrium and only exhibits pairwise interactions. 

We find reasonable agreement between the potential inferred with the GNN and the experimental PMF. Both methods return a measure of the colloidal radius and depletion interaction range that agree with previous characterizations. Additionally, the well depth and curvature of the GNN potential both match the experimental PMF near the potential minimum, indicating that this method could be informative for predicting relative populations of colloidal clusters \cite{meng_free-energy_2010} or the elastic properties of colloidal crystals \cite{meng_elastic_2014}. 

Polydispersity and experimental noise are the likely causes of the disagreement between the two potentials at short distances. The localization algorithm occasionally identifies particles closer than 2 particle radii apart (\SI{1.3}{\micro\meter}), which is expected because \SI{1.3}{\micro\meter} is the average diameter but the standard deviation is approximately 2.5\%.
The small number of samples for center-to-center distances smaller than around \SI{1.2}{\micro\meter} is reflected in the larger uncertainty in the inferred potential in this region. 

As a final verification, we compare the pair correlation function for the experimental data to one generated by running simulations with a representative inferred interaction potential (Fig.~\ref{fig:exp}b). We see that the two pair correlation functions match well, with first and second peaks being at the same pairwise distances and having the same strengths.  In the region \SI{1.6}{\micro\meter} to \SI{1.8}{\micro\meter}, the pair correlation from the simulations has a trough that is absent in the experimental data.  
This observed discrepancy in $g(r)$ may arise from hydrodynamic interactions, which are present in the experiment and will therefore affect the potential inferred from the experimental data, but are not explicitly accounted for in our model.
Specifically, our model does not account for the decreased mobility and increased force required to move particles closer as interparticle distance decreases, a consequence of hydrodynamic interactions~\cite{brenner1963stokes}. This configuration-dependent effect mediated by fluid dynamics in experimental systems is likely born out as an artificial repulsion in the interparticle potential.

An alternative method for inferring effective interaction potentials, called the Iterative Boltzmann Inversion Algorithm~\cite{moore2014derivation,Reith2003_iterative_boltz_inversion}, begins by setting the inferred pair potential to the negative log of the pair correlation function. It then iteratively updates the potential by running simulations and computing differences between the pair correlation function from simulations run with the approximate potential and the true pair correlation function from the experiments. 
While our method does not yield perfect agreement between the simulated pair correlation function (green line in Figure~\ref{fig:exp}) 
 and the experimental pair correlation function (black dots in Figure~\ref{fig:exp}), the significant agreement between the two curves provides further evidence that we have recovered an effective potential that reflects the experimental interactions.

Employing our method provides two key advantages over the iterative Boltzmann Inversion approach to reconstructing the effective pair potential. Firstly, iterative Boltzmann inversion cannot be applied to derive potentials in non-equilibrium settings, whereas our maximum-likelihood-based approach has no such limitations. Secondly, any approaches based on calculating $g(r)$, like iterative Boltzmann Inversion, inherit the uncertainties of the binning procedure used to construct the coarse-grained $g(r)$. Our approach trains on the exact experimental positions of the particles and is therefore more precise.


\section{\label{sec:discussion} Discussion}

We have demonstrated a general method, rooted in physical dynamics, for learning interaction potentials from stochastic trajectory data. We have validated it here on simulated Brownian data, simulated underdamped Langevin dynamical data, and experimental colloidal data. We expect the method to generalize to alternative dynamics. For example, straightforward modifications to the transition probability should allow application to active particle systems and processes displaying non-Gaussian noise statistics~\cite{fodor2018non}. 

Previous work fitting interaction potentials with neural networks has relied on either (1) data labeled with energies, as in the case of fitting to data from DFT simulations~\cite{batzner20223}, (2) deterministic motion, so that matching the deterministic trajectories could serve as a loss function~\cite{lemos2022rediscovering}, (3) highly costly computations, such as full molecular dynamics simulations integrated with the optimization procedure~\cite{wang2023learning}, or (4) assumed forms or bases for the interaction potential~\cite{frishman2020learning}. Our method is well-suited to stochastic data, has an efficient cost function, does not require labeled data, and works for arbitrary functional forms. 

To our knowledge, our work also constitutes the first time a stochastic potential inference method has been validated on bulk experimental data. We have validated our maximum-likelihood approach by reconstructing both simulated and experimental potentials by using a GNN to describe particle interactions. Our method can be applied equally to non-equilibrium data, opening up a new frontier of experimental settings -- including active matter systems, defects in active nematics, cell-cell interactions, and tribocharged particles -- for which interparticle interaction potentials can be inferred. 

There are two primary limitations of our method: the need for an equation of motion for the system, and the need for sufficiently well-sampled data. While many systems of interest, including active matter systems, often have well-described equations of motion, one could also envision interfacing our method with one that learns an equation of motion~\cite{brucknerInferringDynamicsUnderdamped2020b}, a strategy that has been gaining traction in recent years. The second limitation is the need for sufficient sampling: if the data is low-density, the system will not sample close configurations, limiting the accuracy of the inferred potential in that regime. Likewise, if the $\Delta t$ between adjacent frames is too large, it becomes increasingly difficult for the maximum likelihood method to distinguish the deterministic forces from the stochastic noise. However, this is a limitation for all interaction potentials inference methods applied to stochastic data.

It is unknown whether this method could resolve two very different underlying potentials that give rise to the similar dynamics -- for example, differentiating between the interactions for a crystal of attractive particles and a crystal of hard-sphere particles driven to crystallize by entropy. It is possible that the difference in the underlying potentials could be resolved by sampling more particle trajectories, but it is also possible that some prior knowledge would be needed to differentiate them in certain regimes. One potential method for accounting for entropic effects at a finite density is to vary the density and extrapolate to the dilute regime~\cite{iacovella2010pair}, a method that could be integrated with our approach.

Our method to infer interaction potentials will enable measurements of interactions in a much broader range of systems than was previously possible, potentially opening the door to novel discoveries. We anticipate that this data-driven approach to analysis of many-particle systems will become more tractable with computational advancements. Because our setup allows for easily modified neural network architectures, as neural networks become more data-efficient, so will our method. We also anticipate application of our method to material design, as inferred interactions from complex systems would allow us to accurately simulate, and thus design behavior in, otherwise inaccessible materials.

\appendix

\section*{Appendix}

\section{\label{sec:methods}Simulations}
We simulate particle trajectories using JAX-MD \cite{jaxmd2020}. Our system consists of 128 particles interacting via a Morse potential with $\varepsilon=1.0, \alpha=5.0, \sigma=1.0$, with either Brownian or Langevin dynamics. We use a timestep $\Delta t=10^{-5}$, a friction coefficient $\gamma=0.1$, a temperature $k_B T = 1.0$, and a number density of 0.5. We performed tests for both two and three dimensional simulated systems. The figures above show results in three dimensions; we were equally able to recover the interaction potential for simulations of two dimensional systems of particles. We run 50,000 timesteps and save pairs of frames every 500 steps in order to get quasi-independent snapshots, and we repeat this procedure with 500 different random keys that produce different initial configurations, for a total of 50,000 frame pairs used for training: (50,000 timesteps / snapshot every 500 steps) $\times$ 500 configurations = 50,000. The pairs of frames that we save are adjacent, separated in time by $\Delta t = 10^{-5}$. Additional testing has shown that while using quasi-independent pairs of frames reduces the amount of data that we need to fit the potential, we can still infer the interaction potential from frame pairs that are more highly correlated in time. In the case of inferring parameters of a known Morse potential, using only 500 pairs of frames produced results that were nearly as accurate as when using 50,000. When fitting, we sweep over parameter ranges $\varepsilon = [0.5, 1.5], \alpha=[3.0, 7.0], \sigma=[0.5, 1.5]$ in increments of 0.1, 0.27, and 0.1 respectively. Results are independent of choice of range. 

When we compute the pair correlation function $g(r)$, we run a simulation with a representative interaction potential learned from the GNN. Below, we show the interaction potential used in the simulation plotted against an average of 10 potentials learned via the GNN. These 10 potentials were computed with different random seeds used in the training process. The representative potential was chosen to approximate the average inferred potential. We additionally include the experimental PMF data in the comparison.
\begin{figure}[pb]
	\centering
\includegraphics[width=\linewidth]
 {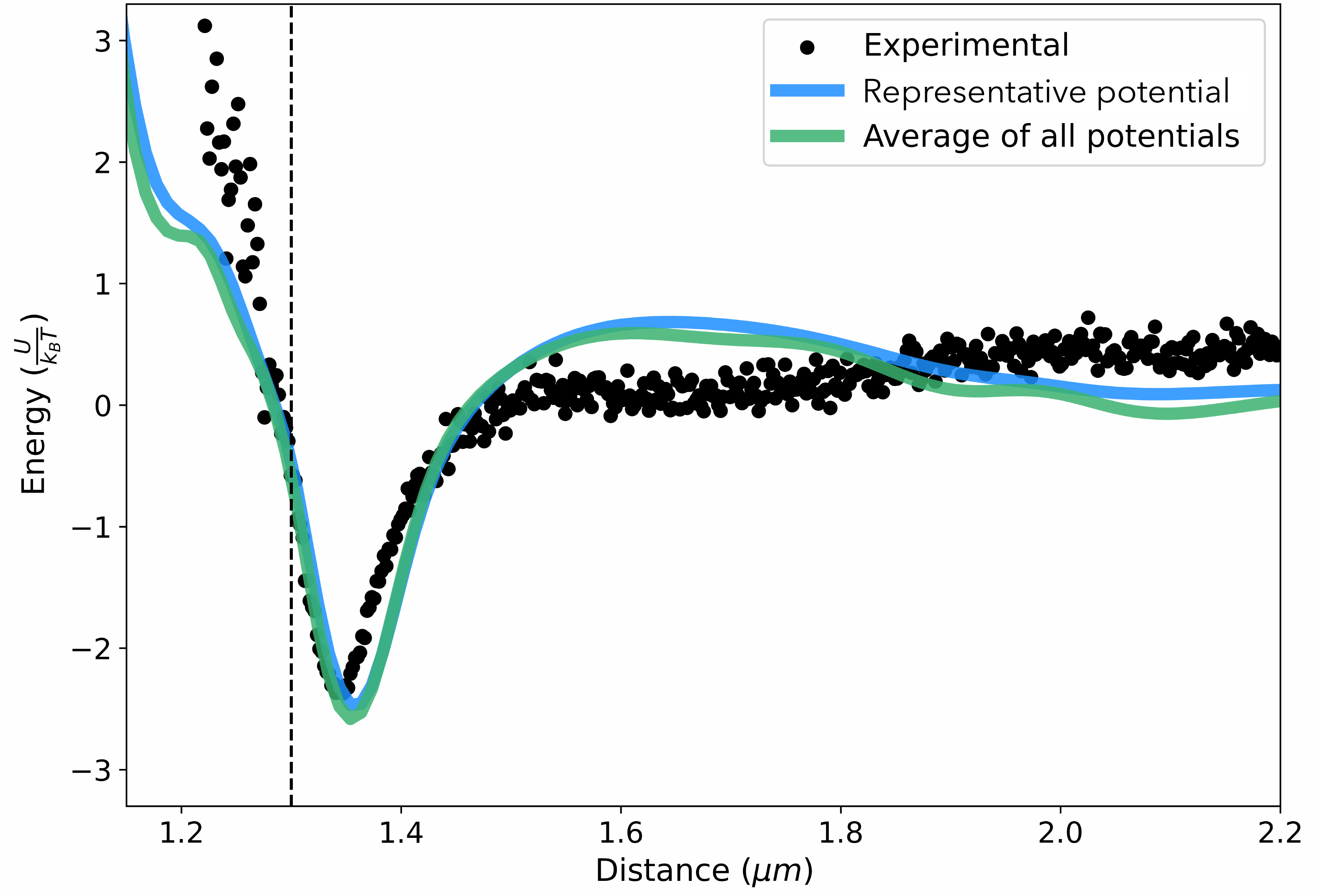}
	\caption{\label{fig:rep_gr} \textbf{Representative vs average  potential.} The blue line shows the interaction potential used to compute $g(r)$ from simulations and the green line shows an average of 10 interaction potentials found via the GNN with different random seeds. The black dots show experimental results. 
 }
\end{figure}

\section{Neural Network}
The neural network we use to infer the interaction potential with free functional form is NequIP, an E(3) equivariant neural network for which the architecture is given in~\cite{batzner20223}. We use the JAX-MD implementation of NequIP for training with the following parameters: a cutoff=6.0 (approximately six times the particle diameter), n\_elements=1, hidden\_irreps=128x0e, sh\_irreps=1x0e, shift=0.0, and scale=1.0. We use the adam optimizer from optax~\cite{deepmind2020jax} with a learning rate of 1e-3. We train it on the simulated Brownian dataset described in section~\ref{sec:brownian} above, using the full 50,000 data points. Surprisingly, we find good agreement between the neural network potential and the true interaction potential after only 10 epochs of training. We also train the network on the 8,340 experimental snapshots described below, obtaining the results in Fig~\ref{fig:exp} after 50 epochs of training.

\section{Experiments}
For the experimental validation, we suspend \SI{1.3}{\micro\metre} charge-stabilized sulfate latex colloidal spheres (Molecular Probes Lot S37499) in deionized water (output from Millipore Elix 3 and Millipore Milli-Q Synthesis) and induce a depletant interaction with carboxymethyl cellulose salt polymers (NaCMC, DS 0.9, $\geq$99.5\%, Acros Organics, molecular weight = 700 \unit{\kilo \dalton}) with a radius of gyration of \SI{60}{\nano\metre} at a concentration of 0.075 \unit{\milli\gram\per\milli\liter}. We plasma clean a glass slide and coverslip for 3 minutes to make the surface hydrophilic. We pipette the solution into a thin, quasi-2D sample chamber with a thickness of approximately 5 \unit{\micro\meter} and seal the chamber with vacuum grease and UV cured glue. We allow the particles to sediment and deplete to the bottom surface of the chamber. We find that the particles are not immobilized and are able to freely diffuse across the surface to interact with each other.

We record bright-field images of these particles using an inverted bright-field microscope (Nikon Eclipse Ti TE2000) with a water-immersion objective and correction collar (Plan Apo VC 60×/1.20 WI, Nikon) and a 1024 × 1024-pixel CMOS monochrome sensor array (PhotonFocus A1024). We record videos at a frame rate of 7.164 ms/frame for 1 second intervals every 30 seconds for 30 minutes, for a total of 60 videos. While the high frame rate means that the data are highly correlated across frames within a single video, we find the 30 second intervals between recording to be sufficient to ensure that each data set is uncorrelated with the previous video.

\section{Analysis of Inferred Potentials}
By relating the width of the well to the range of interaction of the depletion attraction, we implicitly assume that the depletant particles are equilibrating significantly faster than the frame rate. We can estimate the equilibration time scale of the depletants by dividing the square of the radius by the diffusion coefficient, which we compute via the Stokes-Einstein relation: $\tau_D \approx \frac{a_{d}^2}{D}$ with $D = \frac{k_B T}{6\pi\eta a_{d}}$, where $a_{d}$ is the radius of gyration of the depletant particles, $T$ is temperature and $\eta$ is the dynamic viscosity of water. We find that $\tau_D = 9.5\cdot 10^{-4}$ s, which is 7.5 times faster than the frame rate.

\begin{figure}[pb]
	\centering
\includegraphics[width=\linewidth]
 {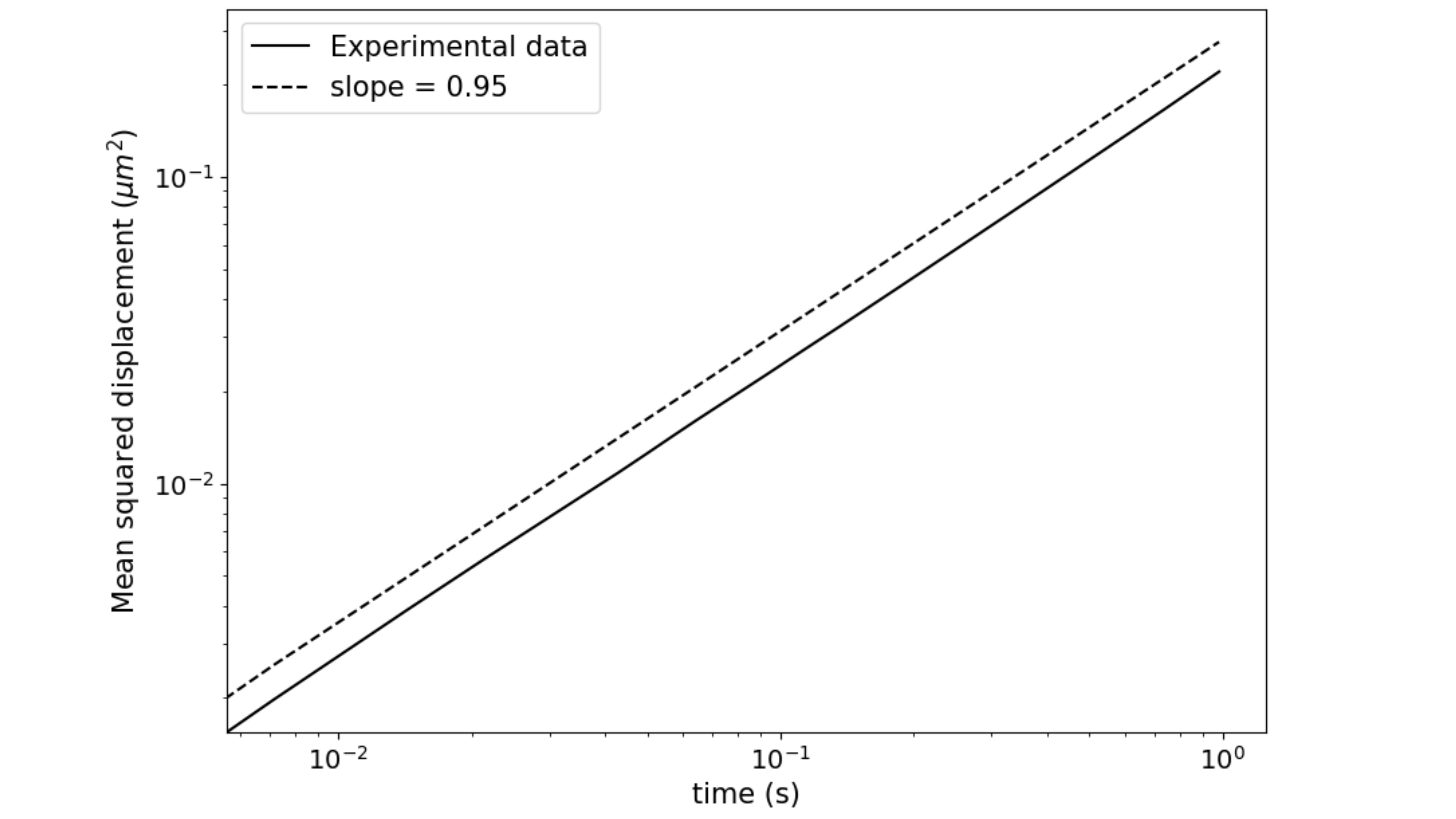}
	\caption{\label{fig:msd} \textbf{Mean squared displacement for the experimental data}. Dashed line shows a slope of 0.95. The experimental data is slightly sub-diffusive, with a slope less than 1.
 }
\end{figure}

\begin{figure}[tpb]
	\centering
\includegraphics[width=\linewidth]
 {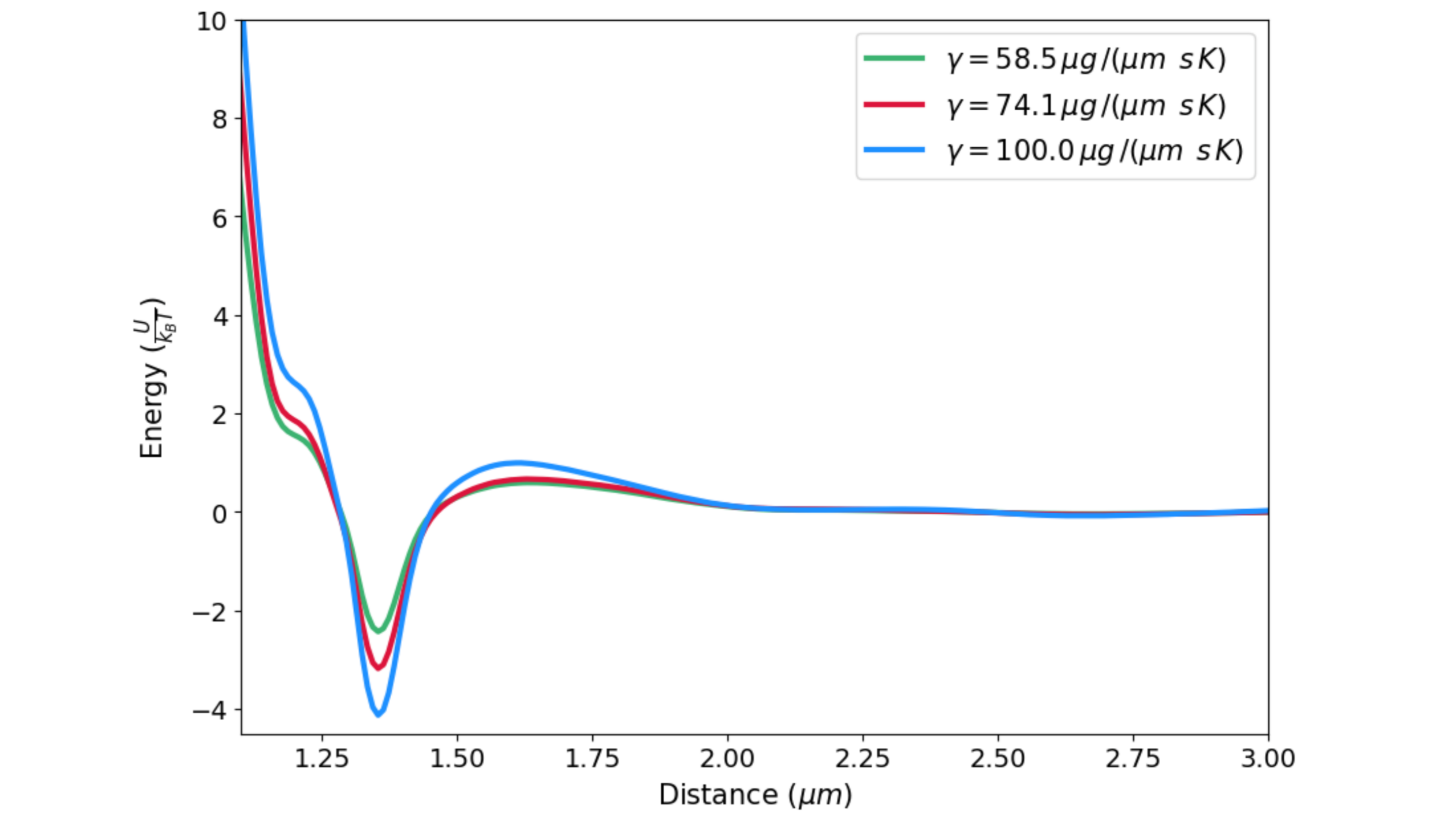}
	\caption{\label{fig:gamma} \textbf{Inferred interaction potential depends on friction coefficient.} The inferred potential is shown for three different choices of the friction coefficient.
 }
\end{figure}

We additionally assume that the friction coefficient is known, or that it can be accurately quantified from the data. If the system is Brownian, the friction coefficient $\gamma$ can be determined by computing the mean squared displacement, $\langle (\Delta x)^2 \rangle = 2 D \Delta t$ where $D$ is the diffusion coefficient and $\Delta t$ is the change in time. We can additionally use the equation $D = \frac{k_B T}{\gamma}$ to extract the friction coefficient $\gamma$. This is the procedure we use to extract the friction coefficient from the experimental data. However, the experimental data is slightly sub-diffusive, as shown in Figure \ref{fig:msd}. If the experimental data were perfectly Brownian, the slope of the curve above would be 1.0, rather than 0.95. The friction coefficient has a significant impact on the inferred potential. If the friction coefficient is larger, more force is needed for the particle to move the same distance, so the potential well is deeper (Fig \ref{fig:gamma}).

\begin{figure}[pb]
	\centering
\includegraphics[width=\linewidth]
 {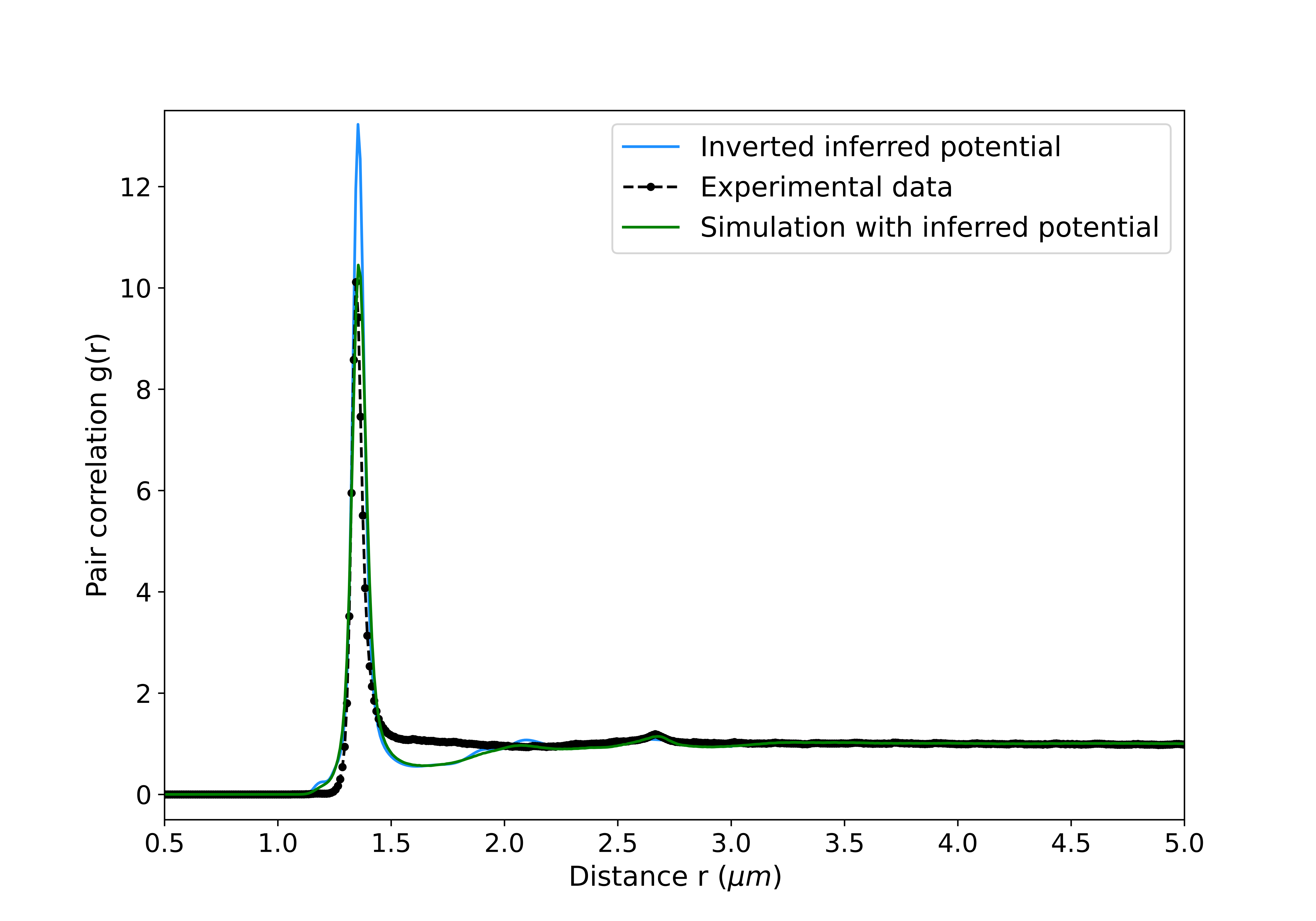}
	\caption{\label{fig:approxgr} \textbf{Pair correlation function estimate} by (green) simulation with inferred potential and (blue) inverted inferred potential.
 }
\end{figure}

Lastly, when we compute the pair correlation function in Fig~\ref{fig:brownian_arbitrary}, we use simulations of particles interacting via a representative inferred potential. That representative potential is given in Fig~\ref{fig:rep_gr}. To check whether the inferred potential might include artifacts that arise from three-body and higher-order correlations, we also compare the measured and simulated pair correlation functions to $g(r) = e^{-\beta U(\textbf{r})}$, where $U(\textbf{r})$ is the inferred potential. The results are given in Fig~\ref{fig:approxgr}. If the inferred potential were to include features arising from higher-order correlations, we would expect the resulting $g(r)$ to have a second peak at $r\approx2\sigma$ or $r\approx\SI{2.6}{\micro\meter}$. Such a peak is seen in both the measured and simulated $g(r)$, where it arises because the system is not in the dilute limit. We do see a small peak at this value in the blue curve of Fig~\ref{fig:approxgr}, but it is not as pronounced as in the measured $g(r)$. Furthermore, there are other small oscillations in the blue curve that result in peaks at other values of $r$. All of these peaks are more likely to arise from overfitting than from higher-order correlations influencing the inferred potential.
 
\bibliographystyle{unsrt}  
\bibliography{biblio_inferring.bib}  

\end{document}